\documentclass[12pt,a4paper]{article}
\usepackage{graphicx}
\usepackage[cp1251]{inputenc}
\usepackage[english]{babel}
\usepackage{amssymb, amsfonts, amsmath}
\usepackage{amscd}
\begin{document}

\date{}

\title{Exact solutions of Maxwell vacuum equations in null homogeneous Petrov spaces with non-solvable motions groups}

\author{V. V. Obukhov}

\maketitle

\noindent
Tomsk State  Pedagogical University, Scientific Department, 60 Kievskaya St., Tomsk, 634041, Russia; \\ \quad

\noindent
Laboratory for Theoretical Cosmology, International Center of Gravity and Cosmos, Tomsk State University of Control Systems and Radio Electronics (TUSUR), 36, Lenin Avenue, Tomsk, 634050, Russia

\quad

\noindent
Correspondence: obukhov@tspu.ru;

\quad

\noindent
Keywords: theory of symmetry, continuous groups of transformations, homogeneous spaces, Maxwell equation.

\section{Introduction}




Homogeneous spaces occupy a special place in the theory of gravity. According to modern concepts, the observable Universe is characterized by large-scale spatial homogeneity and isotropy. Spaces whose symmetry meets this condition include Friedmann, de Sitter, and anti-de Sitter spaces (see \cite{1} - \cite{4}). Most cosmological models are constructed on the basis of these spaces (see, for example, \cite{5} - \cite{17}).

At the same time, a number of researchers propose considering cosmological models in which the requirements for the isotropy of the three-dimensional space-like subspace $V_3$ of the  space-time $V_4$ can be relaxed. Such models include, in particular, models based on space-time manifolds $V_4(N)$, in which the three-parameter groups of motions $G_3(N)$ act simply transitively on the space-like hypersurface $V_3(N)$. These space-time man  are usually called space-time manifolds (or more simple: spaces) of type $N$ according to Bianchi (\cite{18}-\cite{19}, $N$ corresponds to the number of the group $G_3(N)$ in the Bianchi classification). Natural generalizations of spaces of type $V_4(N)$ are spaces in which the three-parameter groups of motions $G_3(N)$ act simply transitively on null hypersurfaces (we will call these spaces  null homogeneous spaces of type $N$), and also spaces $V_4(N)$ with groups of motions $G_3(N)$ acting simply transitively on  time-like hypersurfaces. All spaces with groups of motions $G_3(N)$ acting simply transitively on non-null hypersurfaces we will call non-null homogeneous Petrov spaces of type $N$ and denote as $V_4(N)$. The geometry of a non-null homogeneous Petrov space of type $N$ is completely determined by the geometry of a homogeneous space $V_3(N)$. Under certain conditions (see below), the geometry of null homogeneous spaces of type $N$ is also completely determined by the geometry of a homogeneous space  $V_3(N)$. We will call such spaces null homogeneous Petrov spaces of type $N$ and denote as $V^*_4(N)$. Homogeneous null and non-null Petrov spaces of type $N$ are united into a common class of spaces, which we will call homogeneous Petrov spaces. Homogeneous non-null Petrov spaces have been studied in considerable detail in the theory of gravity, including in cosmology (see, for example, \cite{20}-\cite{24}.

In homogeneous Petrov spaces, the equations of motion of a test particle (the Hamilton-Jacobi and Klein-Gordon-Fock equations) form algebras isomorphic to the  algebras of  the  operators of groups $G_3(N)$. This fact was used in the paper \cite{25} to construct a method for the exact integration of the equations of motion, which the authors called the method of noncommutative integration. This method was later generalized to the case of the Dirac-Fock equation (see \cite{26}-\cite{28}).

The method of non-commutative integration allows one to unite the set of homogeneous Petrov spaces with the set of Stackel spaces (for definitions and an exposition of the theory of the Stackel spaces, see \cite{29} \cite{34}) into a single set of spaces in which the equations of motion of a test particle can be exactly integrated or, in any case, reduced to systems of ordinary differential equations. The method of integration applicable in Stackel spaces is known as the method of complete separation of variables (or as the method of commutative integration. The method of complete separation of variables for diagonal spaces was first applied by Stackel as early as 1897. In honor of Stackel, all spaces in which the method of commutative integration is applicable were called Stackel spaces. The general theory of Stackel spaces was finally constructed in its general form in the works of Shapovalov (\cite{35}-\cite{38}). The method of commutative integration is so called because only in Stackel space can the equations of motion of a test particle admit commutative algebras of linear  operators of symmetry (classical or quantum integrals of motion) that contain momenta in first degree or (and) in second degree. The coefficients before the leading powers are components of Killing vector or tensor fields. These Killing fields, the total number of which (including the metric tensor) coincides with the dimension of the Stackel space, form so called complete sets. The presence of complete sets allows one to use  the complete separation of variables method (in a privileged coordinate system). It is important to note that separation of variables in privileged coordinate systems also takes place in Einstein equations (provided that the matter tensor has the same symmetry as the metric tensor). This circumstance has made Stackel spaces very attractive for research. On the one hand, it  takes
 the efficiently method  obtaining exact solutions of the Einstein and Einstein-Maxwell equations (see \cite{39}-\cite{41}). From the other hand the very structure of the metric tensor of Stackel space and the structure of  the vector-potential of the admissible electromagnetic field (that is, the field in which the classical or quantum equations of motion of a charged test particle admit integration by the complete separation of variables  method ) makes it possible to pose and solve classification problems of finding all nonequivalent exact solutions of the equations of motion or field equations in the corresponding theories of gravitation, as well as in Minkowski space (see, for example, \cite{42}-\cite{45}). Individual solutions obtained due to these classification  are still used to construct realistic models (see, for example, \cite{46}, \cite{46a}).

Stackel spaces are of great importance in the theory of gravity, since essential part of the most interesting from the physical point of view exact solutions of Einstein's equations are Stackel or conformally Stackel spaces. Shapovalov's fundamental theorem proves that a space is Stackel if and only if it admits complete sets of mutually commuting Killing fields. However, this condition is insufficient for the equations of motion of a charged test particle (the Hamilton-Jacobi, Klein-Gordon-Fock, and Dirac-Fock equations) to admit complete separation of variables. In the case of a charged test particle, a necessary and sufficient condition for the applicability of the method of commutative integration is the existence of a commutative algebra of operators of motion. For quantum mechanical equations of motion, this demand take place also  for a neutral test particle (see, \cite{47}).

Many of the listed properties of Stackel spaces are also inherent in homogeneous Petrov spaces. For example, the Einstein and Einstein-Maxwell equations in homogeneous Petrov spaces for an admissible electromagnetic field (invariant under the group $G_3(N)$) also reduce to systems of ordinary differential equations. This occurs because the components of the potential of the invariant electromagnetic field and the components of the metric tensor in the privileged coordinate system contain only functions, each of which depends on only one variable. This allows us to pose and solve the same classification problems for the set of Petrov spaces so as for the set of Stackel spaces.

The first of these classification problems is the classification of all four-dimensional pseudo-Riemannian spaces on whose subspaces groups of motions act simply transitively. These problem was  solved by A.Z. Petrov in the papers \cite{48}-\cite{50}. In paper \cite{51}, the components of the canonical reper vectors (see \cite{52}) were found for all homogeneous Petrov spaces, which made it possible to represent the components of the contravariant metric tensors of a homogeneous Petrov spaces in a form convenient for study.

The second problem. Classification of the potentials of admissible electromagnetic fields for all homogeneous Petrov spaces.  The problem was solved in papers \cite{53} - \cite{56}.

The third problem. After solving the first two classification problems, it became possible to consider the classification of exact solutions to field equations (Maxwell and Einstein-Maxwell equations) for all homogeneous Petrov spaces and admissible electromagnetic fields. In papers \cite{57}, \cite{58} the problem of classification of exact solutions to Maxwell vacuum equations for non null homogeneous Petrov spaces was completely resolved. In papers \cite{59}, \cite{61} the classification problem for electrovacuum spaces and Einstein spaces for non null homogeneous Petrov spaces with an Abelian group of motions was resolved. Some classification problems were considered for Ricci-flat nonzero homogeneous spaces (see \cite{62}, \cite{63}).

A separate area of research is the development of the  non commutative integration method, as well as the study of the symmetry of homogeneous spaces (\cite{64}, \cite{65}). The applicability of the method was investigated, including in Minkowski space (see, for example, \cite{66}), so as for modeling various processes in homogeneous spaces (see \cite{67}).

The goal of this paper is to provide a complete classification of exact solutions of Maxwell vacuum equations in null homogeneous Petrov spaces with unsolvable groups of motions (groups $G_3(VIII), G_3(IX)$).

\quad

 The paper is structured as follows.

\begin{enumerate}

\item The second section provides all necessary definitions and notations used in the text.

\item In the third section the contravariant components of metric tensor of the null homogeneous Petrov spaces with unsolvable groups are represented.

\item In the fourth section the vacuum Maxwell equations are obtained for all null homogeneous Petrov spaces.

\item In the fifth section all nonequivalent exact solutions of Maxwell vacuum equations in null homogeneous Petrov spaces with unsolvable groups are found.

\end{enumerate}

\section{HOMOGENEOUS PETROV SPACES}

Throughout the text, the following notations for indices are used.
$$ i,j,k,l \div 1,2,3,0;\quad p,q \div 2,3; \quad \alpha ,\beta ,\gamma  ... \div 1,2,3;\quad a,b,c\div 1,2,3.$$
The fourth coordinate of the  semi-geodetic coordinate system  $\left\{u^i\right\}$ in the spaces $V_4(N)$, $V^*_4(N)$ is supplied with the index $i=0$ ($u^0$), and the numbering of coordinate indices always starts with one and ends with zero (when indices $i,j,k,l$ are used). Greek letters denote the coordinate indices of the coordinate system $\left\{u^{\alpha} \right\}$, related to the invariant space $G_{3} \left(N\right)$. In addition, Greek letters denote functions of the variable \quad  $u^0$, \quad $\xi, \quad \varepsilon, \quad \varepsilon_a =\pm 1 $, \quad  $\epsilon = 0,\pm 1.$ \quad
The Latin letters $a, b, c$ denote indices of structure constants, tensor indices
of nonholonomic coordinates associated with the canonical frame ($y^\alpha_{(a)}$) or the Killing frame ($x^\alpha_{(a)}$). When the indices $a,b,p,q$ are enclosed in parentheses ((a), (q)), they number vector fields in the canonical reper or the Killing reper. When denoting matrix elements, the superscript numbers the columns, and the subscript numbers the rows.

Consider a three-dimensional Riemannian manifold $V_3$ on which the motions group $G_3(N)$ acts simply transitively ($N$ is the number of the group $G_3(N)$ in the Bianchi classification). According to the definition, in this case the space $V_3$ is called a homogeneous space  $V_3(N)$ ( or an invariant space of the group $G_3(N)$).
The invariant space $V_3(N)$ is associated with two sets of vector fields. One of these sets contains the Killing vector fields $x^\alpha_{(a)}$ and defines generators of infinitesimal motions of the space $V_3(N)$:
\begin{equation}\label{1}
X_a = x_{(a)}^\alpha \mathrm{p}_a, \quad [X_a, X_b]= \mathrm{\bar{C}}^c_{ab}X_c,
\end{equation}
The other set consists of vectors $y^\alpha_{(a)}$, which form the canonical frame.  This set defines generators of finite motions of the space $V_3(N)$:
\begin{equation}\label{2}
Y_a = y_{(a)}^\alpha \mathrm{p}_\alpha , \quad [Y_a, Y_b]= C^c_{ab}Y_c,
\end{equation}
($\mathrm{\bar{C}}^c_{ab}, \quad C^c_{ab}$ are structural constants of the group $G_3(N)$).
As it is known, the metric tensor $\mathrm{{G}}^{\alpha\beta}$ of the space $V_3(N)$ can be represented in the form (see \cite{52}):
\begin{equation}\label{3}
\mathrm{\bar{G}}^{\alpha\beta} = y_{(a)}^\alpha y_{(b)}^\beta \tilde{\eta}^{ab},
\end{equation}
$\tilde{\eta}^{ab}$ are arbitrary constants.
The components of the Killing vector fields of the space $V_3(N)$ satisfy the Killing equations:
\begin{equation} \label{4}
\mathrm{\bar{G}}^{\alpha\beta}_{,\gamma}x_{(a)}^{\gamma} =\mathrm{\bar{G}}^{\alpha\gamma}x^{\beta}_{(a),\gamma} + \mathrm{\bar{G}}^{\beta\gamma}x^{\alpha}_{(a),\gamma},
\end{equation}
The Killing vector fields $x^{\alpha}_a$ are connected with the vector fields of the canonical frame $y_{(a)}^{\alpha}$ by the sets of equations (see \cite{52}):
\begin{equation} \label{5}
y_{(a),\beta }^{\alpha} =x_{\beta}^{(b)} x_{(b),\gamma }^{\alpha}y_{(a)}^{\gamma} \Leftrightarrow
x_{(a),\beta}^{\alpha} =y_\beta^{(b)} y_{(b),\gamma}^{\alpha } x_{(a)}^{\gamma}.
\end{equation}
Here
\begin{equation} \label{6}
x_{\beta }^{(a)} x_{(a)}^{\alpha} = y_{\beta }^{(a)} y_{(a)}^{\alpha} =\delta_\beta^{\alpha}.
\end{equation}
Using the contravariant  metric tensor  \eqref{3} of the homogeneous space $V_3(N)$, one can construct a contravariant metric tensor $g^{ij}$ of pseudo Riemannian homogeneous space $V_4(N)$, which in semigeodesic coordinate systems $[u^i]$  has the form:
\begin{equation} \label{7}
g^{ij}(u^i) =\left(\begin{array}{cccc} {} & {} & {} & {\epsilon_0} \\ {} & {G^{\alpha \beta }} & {} & {0} \\ {} & {} & {} & {0} \\ {\epsilon_0} & {0} & {0} & {\epsilon_1} \end{array}\right), \quad G^{\alpha\beta} = y_{(a)}^\alpha y_{(b)}^\beta \eta^{ab},
\end{equation}
$$
\eta^{ab} = \eta^{ab}(u^0), \quad \left\|g^{ij}\right \| < 0, \quad \epsilon_\varsigma =0, \pm1 \quad (\epsilon_0 \epsilon_1 =0).
$$
In the case where \quad $\epsilon_0=0,$ \quad the space $V_4(N)$ is called non null homogeneous Petrov space of type $N$ (or non null homogeneous space $V_4(N)$ of type $N$ according to Bianchi classification if a homogeneous space $V_3(N)$ has a space-like metric).
In this paper, we consider the case where \quad $\epsilon_0=1 \quad \epsilon_1=0$. If the metric tensor \eqref{7} satisfies the Killing equations:
\begin{equation} \label{8}
g^{ij}_{,\alpha}x_{(a)}^{\alpha} = \delta^j_\beta(g^{i\alpha}x_{(a),\alpha}^{\beta}) + \delta^i_\beta (g^{j\alpha} x_{(a),\alpha}^{\beta})
\end{equation}
and the components of the Killing vector fields don't depend on the wave variable $u^0$:
\begin{equation} \label{9}
x_{(a),0}^\alpha =0,
\end{equation}
the corresponding spaces are called the null homogeneous Petrov spaces of type $V^{*}_4(N)$.
From equations \eqref{8} it follows (see \cite{48}-\cite{50}):
\begin{equation} \label{10}
x^i_{(j)} =\delta^i_\alpha\delta^b_j x^\alpha_{(b)},\quad x^\alpha_{(a),1}=0.
\end{equation}
The condition \eqref{8} admits to represent Killing equations \eqref{7} in the same form as \eqref{4}:
\begin{equation} \label{11}
G^{\alpha\beta}_{,\gamma}x_{(a)}^{\gamma} = G^{\alpha\gamma}x^{\beta}_{(a),\gamma} + G^{\beta\gamma}x^{\alpha}_{(a),\gamma}.
\end{equation}
Then the tensor $G^{ab}$ can be represented in the same manner as the tensor $\tilde{G}^{ab}$ (see \eqref{4}):
$$
G^{\alpha\beta} = y_{(a)}^\alpha y_{(b)}^\beta \eta^{ab}.
$$
This constraint  naturally selects from the set of all Petrov spaces a set of  null homogeneous  Petrov spaces of type $G^*_{4}(N)$. The geometry of such space is completely determined by the geometry of the invariant space of type $V_3(N)$.
In the present paper, the case where the motion groups of the invariant space $V_3(N)$ are the  unsolvable groups (groups $V_3(VIII),$ $ G_3(IX)$) is considered.
\quad

\quad

\quad

\section{ Null homogeneous  Petrov spaces of types $G^*_{4}(VIII),$  $G^*_{4}(IX)$}\label{sec5}

The problem of classification of the pseudo-Riemannian manifolds $V^*_4$ with three-parameter groups of motions $G_3(N)$ acting simply transitively on the null hypersurfaces  was solved in the books (see Petrov~\cite{50} (f. 25.4) p. 157).

Provided that the requirement \eqref{11} is satisfied, this classification contains a complete classification of zero homogeneous Petrov spaces of type $V^*_4(N)$.
In the present work, one will use the same coordinate systems as in Petrov's book (see \cite{50}, formulas (25.35)-(25.37) on p. 166.).
There exist four nonequivalent homogeneous spaces $V_3(N)$ that are invariant under the action of non-solvable three-parameter of motions groups. Three of these spaces are generated by the group $G_3(VIII)$ (them will be denoted as $V^{(\iota)}_3(VIII), \quad \iota=1, 2, 3 $), one space is generated by the group $G_3(IX)$ (it will denoted as $V_3(IX)$).

\subsection{${Group \quad G_{3}(VIII)}$}

The nonzero structure constants of the group $G_{3}(VIII)$ are chosen  in the same form as in cite{Petrov}:
$$
C^1_{12} =1, \quad C^3_{23} =1, \quad C^2_{31} = -2\Rightarrow
$$
\begin{equation}\label{12}
\left[{\rm X} _{1} {\rm X} _{2} \right]={\rm X} _{1}, \quad \left[{\rm X} _{2}{\rm X}_3 \right]={\rm X} _{3}, \quad \left[{\rm X} _{3} {\rm X} _{1} \right]=2{\rm X} _{2}.
\end{equation}
Klling vector fields and the coordinate system $(u^\alpha)$ can be found from the communication relations \eqref{12}.
It is known that any contravariant vector field can be diagonalized by choosing a coordinate system. In our case, there is next constraint on coordinate transformations (see cite{Petrov}, p. 158):
13\begin{equation}\label{13}
\tilde{u}^\alpha = \delta^\alpha_1 u^1 + \Phi^\alpha(u^2,u^3) \quad c^\alpha = const,
\end{equation}
It can be shown that coordinate transformations \eqref{13} allow any operator $X_a$ (without loss of generality, assume that this operator is $X_2$) to be represented in one of two forms:
\begin{enumerate}
\item $X_2= f(u^2u^3)p_1,$
\item $X_2= p_2.$
\end{enumerate}
Since the components of the Killing vector field don't dependent on the variable $u^1$, it is easy to show that in the first case the commutation relations \eqref{12} do not hold. Therefore,  $X_2= p_2.$ From the structural equations \eqref{12} it follows:
\begin{equation}\label{14}
{\rm X}_{1} =\Omega_{1}\exp(-u^{2}), \quad {\rm X}_{3} =\Omega_{2}\exp(u^{2}),
\end{equation}
where
$\Omega_{s} =a^\alpha_{s}(u^3) p_{\alpha}.$
To represent the solutions of the commutation relations \eqref{12} in the following form
\begin{equation}\label{15}
{\rm X} _{3} =\left(p_{1} - 2u^{3} p_{2} +\left({u^{3}}^{2} -\varepsilon \right)\right)\exp u^{2},
\end{equation}
$$
{\rm X} _{2} =p_{2},\quad {\rm X} _{1} =p_{3} \exp \left(-u^{2} \right),
$$
one can use the  transformations of the variables from the equations \eqref{13}:
$$
\tilde{u}^1 = u^1 + \Phi^1(u^3), \quad \tilde{u}^2 = u^2 + \Phi^2(u^3), \quad \tilde{u}^3 = \Phi^3(u^3)
$$
The quantity $\varepsilon$ takes one of three possible values: $\quad \varepsilon = 0, \pm 1.$ \quad Therefore, there exist three nonequivalent invariant spaces of type $V_3(VIII)$, which generate three nonequivalent null homogeneous spaces of Petrov type $V^{*(\varepsilon)}_4(VIII)$.

To fully exploit the symmetry of homogeneous space in gravity theory, it is necessary to find the vector fields of the canonical coordinate system $y^\alpha_{(a)}$ . To do this, it is necessary to find solutions to the system of equations \eqref{5}

Let us represent the of Killing vector fields components as:
\begin{equation}\label{16} \left\{\begin{array}{c}
{x^\alpha_{(a)} = \delta_a^1 \delta_2^\alpha + \delta_a^2(u^2\delta_2^\alpha + \delta_3^\alpha) + \delta_a^3(-\delta_1^\alpha\exp{u^3} +}\\
{+({u^2}^2 +\varepsilon\exp{2u^3}) \delta_2^\alpha + 2u^2 \delta_3^\alpha) \Rightarrow} \\
{x^\alpha_{(a),\beta} = \delta^2_\beta \delta_2^\alpha \delta_a^2 + \delta^3_a (2\delta_\beta^2(u^2\delta_2^\alpha + \delta_3^\alpha)+}\\
{ + \delta_\beta^3( - \delta_1^\alpha \exp{u^3} + 2\varepsilon \delta_2^\alpha \exp{2u^3}).}
\\ { x_\alpha^{(a)} = \delta^1_\alpha(\delta^a_1(\varepsilon\exp{u^3} -{u^2}^2\exp(-u^3)) +  }
\\ + (2\delta_2^a u^2- \delta_3^a) \exp{(-u^3})) + \delta_\alpha^2\delta_1^a - u^2 \delta_\alpha^3 \delta_1^a. \quad \end{array}\right. \end{equation}
Using the set of Equation \eqref{16}, one can obtain the system of equations \eqref{5} in the form:
\begin{equation} \label{17}
\left\{\begin{array}{c} {y_{\left(a\right),1}^{1} =y_{\left(a\right)}^{3}, \quad y_{\left(a\right),1}^{2} =-2\varepsilon y_{\left(a\right)}^{3} \exp u^{3},}\\

{y_{\left(a\right),1}^{3} =-2y_{\left(a\right)}^{2} \exp \left(-u^{3} \right),\quad y_{\left(a\right),3}^{1} =y_{\left(a\right),3}^{3}= 0, } \\

y_{\left(a\right),3}^{2} =y_{\left(a\right)}^{2}, \quad {y_{\left(a\right),2}^{\alpha } =0,} \Rightarrow \\
\end{array}\right.
\end{equation}
\begin{equation} \label{18}
y_{\left(a\right)}^{1} =y^1_a(u^1), \quad y_{\left(a\right)}^{2} =\ell^2_a(u^1)\exp u^3, \quad y_{\left(a\right)}^{3} =\ell^3_a(u^1),
\end{equation}
$\ell^1_a =\ell ^1_a(u^1).$ Let us substitute equations \eqref{18} into \eqref{17}. As a~result, one obtains the following equations:
\begin{equation}\label{19}
\dot{\ell}_{\left(a\right)}^1 =\ell_{\left(a\right)}^{3} \quad \dot{\ell}_{\left(a\right)}^{2} =-2\varepsilon \ell_{\left(a\right)}^{3} \quad \dot{\ell}_{\left(a\right)}^{3} =-2\ell_{\left(a\right)}^{2}
\end{equation}
The dot denotes the derivative of $u^{1}$.
Depending on the values of the parameter $\varepsilon $, one can obtain following solutions of the set of Equation \eqref{19}:

\quad

\noindent
${\bf 1}.~\varepsilon =0$ (space of type $V_3^{(0)}(VIII)$)
$$
\ell_{\left(a\right)}^{2} =-c_{\left(a\right)}^{3}, \quad \ell_{\left(a\right)}^{1} =c_{a}^{3} {u^{1}}^{2} +c_{a}^{2} u^{1} +c_{a}^{1} ,\quad  \ell_{\left(a\right)}^{3} =2c_{a}^{3} u^{1} +c_{a}^{2}.
$$
$c_{a}^{\alpha}=const. \quad $  Matrix $y_{a}^\alpha $ can be represented as follows:
\begin{equation} \label{8.9}
y_{\left(a\right)}^{\alpha } =\left(\begin{array}{ccc} {1} & {0} & {0}\\  u^1 & 0 & 1\\ {u^1}^2 & - \exp u^3 & 2u^1\end{array}\right) \Rightarrow
\end{equation}
\begin{equation} \label{21}
\left\{\begin{array}{l}
{{\rm Y}_{1} ={\rm p}_{1} \quad Y_{2} ={\rm p}_{3} +u^{1} {\rm p}_{1} , \quad {\rm Y}_{3} ={u^{1}}^{2} {\rm p}_{1} +2u^{1} {\rm p}_{3} -{\rm p}_{2} \exp u^{3},}\\
{C^1_{12} =1, \quad C^2_{31} = -2, \quad C^3_{23} =1.}
\end{array}\right.
\end{equation}
Let us transform the operator $X_3: X_3\Rightarrow \frac{1}{2} X_3$. Then \eqref{21} takes the form:
$$
\left\{\begin{array}{l}
{{\rm Y}_{1} ={\rm p}_{1} \quad Y_{2} ={\rm p}_{3} +u^{1} {\rm p}_{1} , \quad {\rm Y}_{3} =\frac{1}{2}({u^{1}}^{2} {\rm p}_{1} +2u^{1} {\rm p}_{3} -{\rm p}_{2} \exp u^{3},})\\
{C^1_{12} =1, \quad C^2_{31} = -1, \quad C^3_{23} =1.}
\end{array}\right.
$$
Matrix  $g^{\alpha \beta } = y_{\left(a\right)}^{\alpha } y_{\left(b\right)}^{\beta} \eta^{ab} $  has the form:
\begin{equation} \label{22}
\left\{\begin{array}{c} {g^{11} =\left(\alpha_{11} +2u^{1} \alpha_{12} +{u^{1}}^{2} \left(\alpha_{22} +2\alpha_{13} \right)+2{u^{1}}^{3} \alpha_{23} +\alpha_{33} {u^{1}}^{4} \right)},
\\ {g^{12} =-\exp u^{3} \left(\alpha_{13} +a_{23} u^{1} +\alpha_{33} {u^{1}}^{2} \right),\quad
g^{22} =\alpha_{22}\exp 2u^{3} },
\\ {g^{33} =\frac{1}{4}(\alpha_{22} +4u^{1} \alpha_{23} +4{u^{1}}^{2} \alpha_{33}) ,\quad
g^{23} =-\frac{1}{2}\exp u^{3} (\left(\alpha_{23} +2u^{1} \alpha_{33} \right)}),
\\ {g^{13} =\frac{1}{2}(\alpha_{12} +u^{1} \left(\alpha_{22} +2\alpha_{13} \right)+3{u^{1}}^{2} \alpha_{23} +2\alpha_{33} {u^{1}}^{3}) } \end{array}\right.
\end{equation}

\quad

\noindent
${\bf 2}.~\varepsilon =-1$ (space of type $V_3^{(-)}(VIII)$)
Then, solutions of Equations \eqref{17}--\eqref{19} can be represented in the following  form:
%
%
%
\begin{equation} \label{23}
y_{\left(a\right)}^{\alpha } =\left(\begin{array}{ccc} {1} & {0} & {0}\\ {\sin u^1} & {\sin u^1 \exp u^3} & {\cos u^1} \\ {\cos u^1} & {\cos u^1\exp u^3} & {-\sin u^1}\end{array}\right) \Rightarrow
\end{equation}
$$
\left\{\begin{array}{l}
{Y_1=p_1, \quad Y_2 =(p_1 +p_2 \exp{u^3})\sin{u^1} + p_3 \cos{u^1},}\\
{ Y_3 =(p_1 +p_2 \exp{u^3})\cos{u^1} - p_3 \sin{u^1} \Rightarrow }\\
{C^3_{12} =1, \quad C^2_{31} =1, \quad C^1_{23} = -1.}
\end{array}\right.
$$
The components of the matrix $g^{\alpha \beta } $ are as follows:
\begin{equation} \label{24}
\left\{\begin{array}{l} {g^{11} =\alpha_{11} +\rho -\varrho\cos(2u^{1}) + \alpha_{23} \sin(2u^{1})} +\\ {+2(\alpha_{12} \sin(u^{1}) + \alpha_{13}\cos(u^{1}) )},
\\ {g^{12} =(\rho-
\alpha_{12} \sin(u^{1}) + \alpha_{13}\cos(u^{1})  + \alpha_{23}\sin(2u^{1})}+ \\ { +\alpha_{12} \sin(u^{1}) +\alpha_{13}\cos(u^{1}) )\exp u^{3} },
\\  {g^{22} =\exp 2u^{3}(\rho -\varrho\cos(2u^{1}) +  \alpha_{23} \sin(2u^{1}))},
\\ {g^{33} =\rho+\varrho\cos(2u^{1}) - \alpha_{23} \sin(2u^{1}) },
\\ {g^{13} =\alpha_{12}\cos(u^{1}) - \alpha_{13} \sin(u^{1}) +\varrho\sin(2u^{1}) +\alpha_{23}\cos(2u^{1}) }
\\ g^{23} =[\varrho\sin(2u^{1}) -\alpha_{23} \cos(2u^{1}) ]\exp u^{3}  \end{array}\right.
\end{equation}

\quad 

\noindent
 ${\bf 3}.~\varepsilon =1$ (space of type $V_3^{(+)}(VIII)$).
Then, solutions of Equations \eqref{17}--\eqref{19} can be represented in the following  form:
\begin{equation}\label{25}
y_{\left(a\right)}^{\alpha } =\left(\begin{array}{ccc} {1} & {0} & {0} \\ {\cosh u^1} & {\cosh u^1\exp u^3} & {\sinh u^1} \\ {\sinh u^1} & {\sinh u^1 \exp u^3} & {\cosh u^1}\end{array}\right) \Rightarrow
\end{equation}

$$
\left\{\begin{array}{l}
{Y_1=p_1, \quad Y_3 =(p_1 +p_2 \exp{u^3})\sinh{u^1} + p_3 \cosh{u^1},}\\
{ Y_2 =(p_1 +p_2 \exp{u^3})\cosh{u^1} + p_3 \sinh{u^1},}\\
{C^3_{12} =1, \quad C^2_{31} =1, \quad C^1_{23} = -1.}
\end{array}\right.
$$
The components of the matrix $g^{\alpha \beta } $ are as follows:
\begin{equation} \label{26}
\left\{\begin{array}{l} {g^{11} =\alpha_{11} + \varrho + \rho\cosh(2u^{1}) + \alpha_{23} \sinh(2u^{1})} +\\ {+2(\alpha_{12} \sinh(u^{1}) + \alpha_{13}\cosh(u^{1}) )},
\\ {g^{12} =(\varrho +\rho\cosh(2u^{1}) + \alpha_{23}\sinh(2u^{1}) +\alpha_{13} \sinh(u^{1})} + \\ { +\alpha_{12}\cosh(u^{1}) )\exp u^{3} },
\\
  {g^{22} =\exp 2u^{3}(\varrho + \rho\cosh(2u^{1}) + \alpha_{23} \sinh(2u^{1}))}, \\
 {g^{33} =\rho\cosh(2u^{1}) + \alpha_{23} \sinh(2u^{1}) -\varrho }, \\
  {g^{13} =\alpha_{12}\cosh(u^{1}) + \alpha_{13} \sinh(u^{1}) + \rho\sin(2u^{1}) +\alpha_{23}\cosh(2u^{1}) } \\
 g^{23} =(\rho\sinh(2u^{1})+\alpha_{23} \cosh(2u^{1}) )\exp u^{3}  \end{array}\right.
\end{equation}


\subsection{${Group \quad G_{3}(IX)}$}

The structural equations have the following form:
\begin{equation}\label{27}
\begin{array}{ccc} {\left[{\rm X} _{1} {\rm X} _{2} \right]={\rm X} _{3} ,} & {\left[{\rm X} _{2} {\rm X} _{3} \right]={\rm X} _{1} ,} & {\left[{\rm X} _{3} {\rm X} _{1} \right]={\rm X} _{2} } \end{array}.
\end{equation}
For diagonalization, one can choose vector $x_{\left(1\right)}^{\alpha }$. Obviously, without loss of generality, the diagonalized vector can be represented as: $${\rm X} _{1} =p_{2}. $$ From the structural equations \eqref{27} it follows:
\begin{equation} \label{28}
\begin{array}{cc} {{\rm X} _{2,2} ={\rm X} _{3} ,} & {{\rm X} _{3,2} =-{\rm X} _{2} \Rightarrow {\rm X}_{2,22} +{\rm X} _{2} =0}. \end{array}
\end{equation}
The solution can be presented in the form:
\begin{equation}\label{29}
{\rm X}_{2} =\Omega_{2}\sin(-u^{2}), \quad {\rm X}_{3} =\Omega_{2}\cos(u^{2}),
\end{equation}
where
$\Omega_{p} =a^\alpha_{p}(u^3) p_{\alpha}.$
Using the  transformations of the variables:
$$
\tilde{u}^1 = u^1 + \Phi^1(u^3), \quad \tilde{u}^2 = u^2 + \Phi^2(u^3), \quad \tilde{u}^3 = \Phi^3(u^3)
$$
one can obtain the solution of the system of equation 27\eqref{9.1} in the form:
\begin{equation} \label{30}
\begin{array}{ccc} {{\rm X} _{1} =p_{2} ,} & {{\rm X} _{2} =p_{3} \cos u_{2} +\frac{\sin u^{2} }{\cos u_{3} } \left(p_{1} +p_{2} \sin u^{3} \right),} & {{\rm X} _{3} ={\rm X} _{2,2} } \end{array}.
\end{equation}
This form corresponds to Petrov~\cite{16} (f. (25.6), p. 157).
Let us find the solution of the system of Equation~(14).
To do this, we use the following  matrices:
The components of Killing vector fields can be represented as:
$$
x_{(a)}^{\alpha} =\delta^1_a\left(\frac{\sin u^2}{\cos u^3}(\delta^\alpha_1 + \delta^\alpha_2 \sin{u^3})  +  \delta_3^\alpha \cos u^2\right) +
$$
$$+\delta^3_a\left(\frac{\cos u^2}{\cos u^3}(\delta^\alpha_1 + \delta^\alpha_2 \sin{u^3})  -  \delta_3^\alpha \sin u^2\right)
$$
$$
x_\alpha^{(a)} = \delta^1_\alpha(\delta^a_1\sin{u^2}\cos{u^3} -\delta_2^a \sin u^3 + \delta_3^a \cos{u^2}\cos{u^3}) + \delta_\alpha^2\delta_2^a +
$$
$$ +\delta_\alpha^3 (\delta_1^a \cos{u^2} - \delta_3^a \sin{u^2}).
$$
$$
x_{(a),\beta}^{\alpha} =\frac{1}{\cos{u^3}}(\delta^2_\beta \delta^3_a (\delta^\alpha_1 + \delta^\alpha_2 \sin{u^3} )  +  \delta^3_\beta\delta_a^1( \delta^\alpha_1\sin{u^3}  + \delta^\alpha_2))
$$
The system of Equation~\eqref{5} will take the following  form:
\begin{equation} \label{31}
\left\{\begin{array}{lll} {y_{\left(a\right),3}^{1} =\frac{y_{\left(a\right)}^{2} }{\cos u^{3} } } & {y_{\left(a\right),3}^{2} =y_{\left(a\right)}^{2} \frac{\sin u^{3}}{\cos u^3} } & {y_{\left(a\right),3}^{3} =0 } \\ {y_{\left(a\right),1}^{1} =y_{\left(a\right)}^{3} \frac{\sin u^{3}}{\cos u^3} } & {y_{\left(a\right),1}^{2} =\frac{y_{\left(a\right)}^{3} }{\cos u^{3} } } & {y_{\left(a\right),1}^{3} =-\cos u^{3}y_{\left(a\right)}^{2} } \end{array}\right.
\end{equation}
Let us obtain the solution of the system \eqref{31}:
$$
y_{\left(a\right)}^{\alpha } = \delta^\alpha_1(\delta^1_a + \frac{\sin u^3}{\cos u^3}(\delta^2_a \sin u^1 + \delta^3_a\cos{u^1}))+
$$
$$+\frac{\delta^\alpha_2 }{\cos u^3}(\delta^2_a \sin u^1 + \delta^3_a\cos{u^1})+\delta^\alpha_3(\delta^2_a \cos{u^1} -\delta^3_a\sin u^1)
\Rightarrow
$$

\begin{equation} \label{32}
y_{\left(a\right)}^{\alpha } =\left(\begin{array}{ccc} {1} & {0} & {0}\\ {\sin u^1\tan u^3} & {\frac{\sin u^1}{\cos u^3} } & {\cos u^1} \\ -{\cos u^1}\tan u^3 & {\frac{\cos u^1}{\cos u^3} } & {\sin u^1}\end{array}\right) \Rightarrow
\end{equation}
$$
\left\{\begin{array}{l}
{{\rm Y}_{1} ={\rm p}_{1}, \quad Y_{2} =\frac{\sin u^1}{\cos u^3}({\rm p}_{2} +{\rm p}_{1}\sin u^3) +{\rm p}_{3}\cos{u^1},}\\
{ Y_{3} =\frac{\cos u^1}{\cos u^3}({\rm p}_{2} - {\rm p}_{1}\sin u^3) +{\rm p}_{3}\sin{u^1}; \quad C^{aa}=1.}
\end{array}\right.
$$
The components of the matrix $g^{\alpha \beta } $ are as follows:
\begin{equation} \label{33}
\left\{\begin{array}{l} {g^{11} =a_{11} +
2\tan u^3(a_{12}\sin u^1+ a_{13}\cos
u^1)} +

\\ {+ (\tan u^3)^2 \left(\rho  -\varrho\cos 2u^1+a_{23} \sin2u^{1}\right)},

\\ {g^{12} =(\frac{1}{\cos u^3})(a_{12}\sin u^1+ a_{13}\cos u^1) +}

\\ {+\tan u^3 (\rho -\varrho\cos 2u^1+a_{23} \sin2u^{1})),}

\\ {g^{22} =(\frac{1}{\cos u^3})^2 (\rho -\varrho\cos 2u^1+a_{23} \sin2u^{1}),}

\\ {g^{33} =(\rho + \varrho\cos 2u^1-a_{23} \sin2u^{1})},

\\ {g^{23} =(\frac{1}{\cos u^3})(\varrho\sin 2u^1+a_{23} \cos2u^{1} )},

\\ {g^{13} =a_{12}\cos u^1 - a_{13}\sin u^1 +}

\\ {+ \tan u^3 (\varrho\sin 2u^1+a_{23} \cos2u^{1})}   \end{array}\right.
\end{equation}
The following notations are used above:
$$
\eta^{ab}=\alpha_{ab}, \quad 2\rho= \alpha_{22}+\alpha_{33}, \quad 2\varrho = \alpha_{22} - \alpha_{33}.
$$

\section{Maxwell vacuum equations}

\subsection{Invariant electromagnetic field }
Consider the Hamilton-Jacobi equation for a charged test particle moving in the pseudo Riemannian space $V_4$ with simply transitive groups of motions $G_3(N) $ in the presence of an external electromagnetic field with potential $A_i$:
\begin{equation}\label{34}
\mathrm{H }= g^{ij}\mathrm{P}_i\mathrm{P}_j=m, \quad \mathrm{P}_i=\mathrm{p}_i+A_i,\quad \mathrm{p}_i=\partial_i\varphi, \quad A_0=0.
\end{equation}
As is known, the integrals of motion of the free Hamilton-Jacobi equation are defined by the Killing vector fields $x^i_{(a)}$ and have the form
\begin{equation}\label{35}
\tilde{X}_a=x_{(a)}^i \mathrm{p}_i.
\end{equation}
It can be shown that equation \eqref{34} admits three first-order independent   integrals of motion of the form \eqref{35} if the electromagnetic field with potential $A_i= \delta^\alpha_i A_\alpha$ (the component $A_0$ is made zero by the gradient transformation of the potential) satisfies the system of equations
\begin{equation}\label{36}
x_{(a)}^\alpha(x^\beta_{(b)}A_\beta)_{,\alpha}= \mathrm{\bar{C}}^c_{ba}x^\alpha _{(c)} A_{(a)},
\end{equation}
$\mathrm{\bar{C}}^c_{ba} $ are the structure constants of the group $G_r$.

Indeed, if the functions $\tilde{X}_a$ are integrals of motion, then they commute with the Hamiltonian function $\mathrm{H}$ with respect to the Poisson brackets:
\begin{equation}\label{37}
\frac{\partial \mathrm{H}}{\partial\mathrm{p}_i}\frac{\partial \tilde{X}_a}{\partial u^i} - \frac{\partial \mathrm{H}}{\partial u^i}\frac{\partial \tilde{X}_a}{\partial \mathrm{p}_i}=
2g^{ik}(x^{j}_{(a)} F_{ji}+(x_{(a)}^l A_l)_{,i}\mathrm{P}_k)=0, \quad F_{ji} = A_{i,j}-A_{j,i}.
\end{equation}
Substituting the functions $\mathrm{H} \quad $ from equation \eqref{34} into equation \eqref{37}, we obtain the equation:
\begin{equation}\label{38}
(2g^{ik}x^k_{(a),j} - g^{ij}_{,k}x^k_{(a)})\mathrm{P}_i \mathrm{P}_j - 2(g^{ik}A_jx^j_{(a),k} + g^{ij}A_{j,k}x^k_{(a)})\mathrm{P}_i = 0.
\end{equation}
Equality\eqref{38} must be satisfied for any independent values of the functions $\mathrm{P}_i$. Therefore, from the equations \eqref{38} it follows:
\begin{equation}\label{39}
g^{ik}x^k_{(a),j} + g^{jk}x^k_{(a),j} - g^{ij}_{,k}x^k_{(a)}=0,
\end{equation}
\begin{equation}\label{40}
g^{ik}A_jx^j_{(a),k} + g^{ij}A_{j,k}x^k_{(a)} = 0 \quad \Rightarrow A_jx^j_{(a),i} + A_{i,j}x^j_{(a)} = 0,
\end{equation}
Equations \eqref{39} are the Killing equations and are satisfied because  $G_3$ is the group of motions of the space $V_4$.
Let us denote:
$$\mathrm{\tilde{A}}_{(a)} =x_{(a)}^\alpha A_\alpha, \quad \mathrm{A}_{(a)} =y_{(a)}^\alpha A_\alpha, \quad /a =x_{(a)}^\alpha\frac{\partial }{\partial u^\alpha}, \quad |a =y_{(a)}^\alpha\frac{\partial }{\partial u^\alpha}. \quad
$$
Then the equations \eqref{40} can be represented as:
\begin{equation}\label{41}
x^i_b (x_{(a)}^j A_j)_{,i} = \mathrm{\tilde{A}}_{a/b} = \mathrm{\tilde{A}}_{a/b} -\mathrm{\tilde{A}}_{b/a} +(x^i_{b/a}- x^i_{a/b})A_i.
\end{equation}
Hence:
\begin{equation}\label{42}
\mathrm{\tilde{A}}_{(a)/b} = \mathrm{\bar{C}}^c_{ba}\mathrm{\tilde{A}}_{(c)},
\end{equation}
which is equivalent to equations \eqref{36}. An electromagnetic field whose vector potential components satisfy condition \eqref{42} is called an admissible electromagnetic field. We will also call it an electromagnetic field invariant with respect to the group $G_r(N)$ (or more simple - invariant electromagnetic field). It can be shown that the vector potential $A_i$ of the invariant electromagnetic field  can be represented in the form:
\begin{equation}\label{43}
A_i = \delta_i^{\alpha}y^{(a)}_\alpha \alpha_{a}, \quad (\mathrm{A}_{(a)}=\alpha(u^0)).
\end{equation}

Consider the problem of existence of a symmetry algebra for the Klein-Gordon-Fock equation in the space $V_n$ with the motions group $G_3$:
\begin{equation}\label{44}
\hat{\mathrm{H}}\varphi=(g^{ij}\hat{\mathrm{P}}_i\hat{\mathrm{P}}_j)\varphi = m^2\varphi, \quad \hat{\mathrm{P}}_j = -\imath \hat{\nabla}_i + A_i.
\end{equation}

Here $\hat{\nabla}_i$ is the covariant operator of the derivative whose connection is consistent with the metric.
The operator $\hat{\nabla}_i$ corresponds to the partial derivative operator  $\hat{\partial}_i =\imath \hat{\mathrm{p}}_i$ with respect to the coordinate \quad $u^i, \Phi$ is the field of a scalar particle with mass $m.$
Let us denote the operator:
$$
{\mathrm{H}}_0=g^{ij} \bar{\nabla} \bar{\nabla}_i.
$$
Then the operator \quad $\hat{\mathrm{H}}$ can be represented as:
\begin{equation}\label{45}
\hat{\mathrm{H}} = -\hat{\mathrm{H}}_0 - \imath{\hat{\mathrm{H}}}_e + A^i A^i, \quad \hat{\mathrm{H}}_0 = g^{ij} \hat{\partial}_i \hat{\partial}_j + \frac{1}{2}(2g^{ik}_i- g^{ik}g^{lm}_{,i}g_{lm})\hat{\partial}_k,
\end{equation}
$$
\hat{\mathrm{H}}_e = 2A^i\hat{\partial}_i + g^{ij} A_{i,j}+ \frac{1}{2}(2g^{ik}_i- g^{ik}g^{lm}_{,i}g_{lm})A_k,
$$
In the works \cite{53}, \cite{55}, \cite{56} it was established that in the space $V_n(r)$ with a simple transitive group of motions $G_{r}$, the equation \eqref{44} admits an algebra of symmetry operators consisting of $r $ independent first-order operators if and only if the electromagnetic field is invariant under the group $G_r$.

We find the commutator of the Hamilton operators \eqref{44} and the operators $\hat{X}_a = \imath x_{(a)}^i \mathrm{p}_i$:
\begin{equation}\label{46}
-[\hat{\mathrm{\mathrm{H}}},\hat{X}_a] = [\hat{\mathrm{H}}_0,\hat{X}_a] + \imath[\hat{\mathrm{H}}_e,\hat{X}_a] - [A^iA_i,\hat{X}_\alpha].
\end{equation}
From the fact that the commutator \eqref{46} is equal to zero, the following system of equations follows:
\begin{equation}\label{47}
g^{ik}x^k_{(a),j} + g^{jk}x^k_{(a),j} - g^{ij}_{,k}x^k_{(a)}=0,
\end{equation}
\begin{equation}\label{48}
2g^{ij}x^j _{(a),ij} + (2g^{ik}_i- g^{ik}g^{lm}_{,i}g_{lm})x^j_{(a),k} -(2g^{jk}_k- g^{jk}g^{lm}_{,k}g_{lm})_{,i}x^i_{(a)} =0,
\end{equation}
\begin{equation}\label{49}
A^i x^k_{(a),i}-x^i_{(a)} A^k{,i} =0, \quad x_{(a)}^k ( g^{ij}A_jA_{i})_{,k} =0,
\end{equation}
\begin{equation}\label{50}
(2g^{ik}A_{i,k} + (2g^{ik}_i- g^{ik}g^{lm}_{,i}g_{lm})A_{k})_{,j}x^j_{(a)}=0.
\end{equation}
Equations \eqref{47} as well as equations \eqref{39} are Killing equations.
From equations \eqref{49} it follows:
\begin{equation}\label{51}
(\hat{\mathbf{A}}_a){,k}+x^l_{(a)} A_{k,l} -x^l_{(a)} A_{l,k} =0,
\end{equation}
which coincides with equations \eqref{36}. Equations \eqref{48}, \eqref{50} are consequences of equations \eqref{39} \eqref{51}.

To check this, it is necessary to use equations \eqref{39} \eqref{51} and their differential consequences:
\begin{equation}\label{52}
g^{ij}_{,ij}x^k_{(a)}= g^{ij}x^k_{(a),ij} + g^{kj}x^i_{(a),ij} + g^{ij}_ix^k_{(a),j},\end{equation}
$$ (g^{ij}A_{ij})_{,k}x^k_{(a)} = -g^{ij}A_{k}x^k_{(a),ij}.$$
Substituting equations \eqref{52} into equations \eqref{48}, \eqref{50}, one obtains the identity.

\subsection{Maxwell vacuum equations}

\quad

Let us consider the Maxwell vacuum equations for invariant electromagnetic field in the null homogeneous Petrov space of the types $V^*_4(VIII)$ and  $V^*_4(IX)$:
\begin{equation}\label{53}
\frac{1}{\sqrt{|g|}}(\sqrt{|g|} F^{ij})_{j}=0 \quad (g=-det|g_{kl}|) \quad \Rightarrow F^{ij}_{,j} -\frac{1}{2g} g_{,j} F^{ij} =0.
\end{equation}
From equations \eqref{21} -- \eqref{32} it follows:
\begin{equation}\label{54}
y_{(1)}^\alpha =\delta^1_a, \quad y_1^{(a)} =\delta_1^a.
\end{equation}
Using \eqref{54} one can represent components $g^{pq}$ in the form:
\begin{equation}\label{55}
g^{pq} =G^{pq} = y^{p}_{(p')} y^{q}_{(q')}\eta^{p'q'}(u^0), \quad G_{pq} = y^{(p')}_{p} y^{(q')}_{q}\eta_{p'q'}, \quad\end{equation} $$G^{pq'}G_{qq'}= \eta^{pq'}\eta_{qq'} = y^{(p')}_{q} y_{(p')}^{p}=\delta^p_q.$$
Due to account \eqref{54} one has:
$$
\det|g^{ij}| = -\det|G^{pq}| =-\exp2(\sigma_0\sigma).
$$
Here $$\exp\sigma = \det|e^{p}_{(q)}|, \quad \exp2\sigma_0 = \det|\eta^{pq}|.$$
The commutation relations can be represented in the form:
\begin{equation}\label{56}
y^\alpha_{(a)|b}-y^\alpha_{(b)|a}=C^c_{ba}y^\alpha_c.
\end{equation}
Let us show that:
\begin{equation}\label{57}
y^\alpha_{(a),\alpha}+\frac{g_{|a}}{2g}=c_a \quad (c_a = C_{pa}^{p}).
\end{equation}
Indeed, from the equality (see \cite{52})
\begin{equation}\label{58}
\frac{g_{|a}}{g}=-g_{ij}g^{ij}_{|a}.
\end{equation}
it follows:
$$
\frac{g_{|a}}{2g}=-\frac{G_{|a}}{2G}= -y^{q}_{(p)|a}y_{q}^{(p)}
$$
$(G=\det{G^{pq}}).$
Using this correlation one can find:
$$
y^\alpha_{(a),\alpha} +\frac{g_{|a}}{2g}= y^{(b)}_\beta y^\beta_{(a)|b} -y^{q}_{(p)|a}y_{q}^{(p)} =
$$
$$
 =y^{(p)}_q y^q_{(p)|a} + C_{pa}^{p} - y^{q}_{(p)|a}y_{q}^{(p)}
 = C_a \quad (C_a = C_{pa}^{p} ).
$$
To obtain Maxwell equations in the null homogeneous Petrov space in a form convenient for use, it is necessary to find the nonholonomic components of the vector:
\begin{equation}\label{59}
\mathrm{F}^a=y_i^{(a)} \nabla_j F^{ij}.
\end{equation}
To make so it is necessary to  use the relations \eqref{54}, \eqref{55}, \eqref{57}.
Nonholonomic components $\mathrm{F}_{ab}, \mathrm{F}_{0b}$ have following  form:
$$
\mathrm{F}_{ab}=y_a^\alpha y_b^\beta F_{\alpha\beta} =C^c_{ba}\alpha_c, \quad \mathrm{F}_{0a}=y_a^\alpha F_{0\alpha}= \dot{\alpha}_a.
$$
The indices $a,b$ are raised and lowered using the nonholonomic components of the metric tensor \quad $\eta^{ab}, \eta_{ac}, \quad (\eta_{ac}\eta^{ac} =\delta^b_a)$:
\begin{equation}\label{60}
\mathrm{F}^{ab}=\eta^{aa_1}\eta^{bb_1}\mathrm{F}_{a_1b_1}.
\end{equation}
Let us introduce the functions $f^a$:
$$
f^a=-(\delta^a_1 \mathrm{F}_{23} + \delta^a_2F_{31} + \delta^a_3 \mathrm{F}_{12}) = (\delta^a_1 C^c_{23} + \delta^a_2 C^c_{31} + \delta^a_3 C^c_{23})\alpha_c,
$$
Then correlation \eqref{60} takes the form:
\begin{equation}\label{61}
\mathrm{F}^{b_1 a_1}=\eta^{aa_1}\eta^{bb_1}(\delta_{[a}^2\delta_{b]}^3 C^{c 1}+\delta_{[a}^3\delta_{b]}^1 C^{c 2}+
\delta_{[a}^1\delta_{b]}^2 C^{c 3})\alpha_c,
\end{equation}
where it is denoted:
$$ C^{c1}=  C^{c}_{23}, \quad  C^{c2}=  C^{c}_{31}, \quad  C^{c3}=  C^{c}_{12}, \quad \delta_{[a}^p\delta_{b]}^q = \delta_a^p \delta_b^q -\delta_b^p\delta_a^q.
$$
Using all these correlations one can find Maxwell equations.

\begin{enumerate}

\item $F^0 = \frac{1}{\sqrt{|g|}}(\sqrt{|g|}F^{0i})_{,0} =0.$

The equation takes following form:
$$
 F^{01}_{,1}- F^{01} \sigma_{,1} = -F_{01,1}+ F_{01} \sigma_{,1} + F^{\quad a}_{1}(y^\alpha_{{(a),\alpha}} - \sigma_{|a}).
$$
From correlations \quad $y_{(1)}^\alpha =\delta^\alpha_1, \quad y^{(1)}_\alpha =\delta_\alpha^1 \quad$ it follows:
\begin{equation}\label{62}
 \alpha_1(u^0) = \mathrm{A}_{(1)}=y_{(1)}^a A_{a}=A_1, \quad \sigma_1 = \sigma_{|1}.
\end{equation}
Then one has:
$F_{01,1}-F_{01} \sigma_{,1} = -\dot{\alpha}_1\sigma_{|1}= -\dot{\alpha}_1 C_{1}.$
From here the equation takes the form:
\begin{equation}\label{63}
C_1\dot{\alpha}_1 = C^c_{b1}\eta^{ab}\alpha_c C_a.
\end{equation}

\item $F^\alpha = \frac{1}{\sqrt{|g|}}(\sqrt{|g|}F^{\alpha i})_{,I} =0. \Rightarrow$
\begin{equation}\label{64}
F^{\alpha i}_{,i}- \sigma_{,i}F^{\alpha i} = F^{\alpha}_{. 1,0} - \dot{\sigma}_{0}F^{\alpha}_{. 1} + F^{\alpha \beta}_{,\beta} - \sigma_{,\beta}F^{\alpha \beta}.
\end{equation}
\end{enumerate}
From \eqref{64} it follows:
$$
y_\alpha^c(F^{\alpha i}_{,i}- \sigma_{,i}F^{\alpha i})= \delta^c_1 (\dot{F}_{0 1} - \dot{\sigma_0} F_{0 1}) + (\dot{\mathrm{F}}^{c}_{\quad 1}- \dot{\sigma}_{0}\mathrm{F}^{c}_{\quad 1})+
$$
$$+ \delta^c_1(y^\beta_{(b),\beta} - \sigma_{|b})\mathrm{F}^{\quad b}_{0}-\mathrm{F}^{ \quad a}_{0}y^{(c)}_\beta y^\beta_{(a),1} +
$$
$$
+ \sigma_{,1}\mathrm{F}^{ \quad c}_{0}
 + \mathrm{F}^{ab}(y^{(c)}_\alpha y^\alpha_{(a)|b}) + F^{cb}(y^\beta_{(b),\beta} - \sigma_{|b})=
$$
\begin{equation}\label{65}
\exp{\sigma_0}((\delta^c_1\dot{\alpha}_1 + \mathrm{F}^{c}_{ \quad 1})\exp{-\sigma_0})_{,0} + \delta^c_1 c_a \mathrm{F}^{ \quad a}_{0} - C_1\mathrm{F}^{ \quad c}_{0} -
$$
$$
 - C^c_{1b}\mathrm{F}^{ \quad b}_{0} +  C_b\mathrm{F}^{cb} + \frac{1}{2}C^c_{ba}\mathrm{F}^{ab}=0.
\end{equation}
Let us denote:
\begin{equation}\label{66}
\omega^{pq}=\eta^{pq}\exp{(-\sigma_0)} \quad (\det{\omega^{pq}}=1), \quad \omega^\alpha= \eta^{1\alpha},
\end{equation}

$$
\alpha =A=(\dot{\alpha}_1 + \omega^2 f^3 -\omega^3 f^2 )\exp{-\sigma_0}, \quad \gamma= Y = \omega^{33} f^2 - \omega^{23} f^3,
$$
$$
\beta=W = \omega^{23} f^2 - \omega^{22} f^3.
$$
Then using \eqref{64} - \eqref{66} one can obtain the Maxwell equations in the form:

\begin{enumerate}
\item
$F^1 = 0:$
\begin{equation}\label{67}
\dot{\alpha} +C^{21}(\alpha \omega^3 +\omega^1 \gamma+f^1(\omega^3 \omega^{23} - \omega^2 \omega^{33}) ) -
C^{31}(\alpha \omega_2 +\omega_1 \beta+
\end{equation}
$$
+f^1(\omega^3 \omega^{22} - \omega^2 \omega^{23}))+ C^{11}(f^1 \exp{\sigma_0}+ \omega^3\beta - \omega^2 \gamma)=0;
$$
\item
$F^2 = 0:$
\begin{equation}\label{68}
C^{22}(\alpha \omega^3 +\omega^1 \gamma+f^1(\omega^3 \omega^{23} - \omega^2\varpi^{33}) +\dot{\alpha}_p \omega^{p3}) -
\end{equation}
$$
 -C^{32}(\alpha \omega^2 +\omega^1 \beta +f^1(\omega^3  \omega^{22} - \omega^2  \omega^{23}) +\dot{\alpha}_p  \omega^{p2})
$$
$$
+ C^{12}(f^1 \exp{\sigma_0}+ \omega^3 \beta - \omega^2\gamma) -\dot{\beta}=0;
$$
\item
$F^3 = 0:$
\begin{equation}\label{69}
C^{23}(\alpha \omega^3 +\omega^1 \gamma+f^1(\omega^3  \omega^{23} - \omega^2 \omega^{33}) +\dot{\alpha}_p   \omega^{p3}) -
\end{equation}
$$
-C^{33}(\alpha \omega^2 +\omega^1 \beta+f^1(\omega^3 \omega^{22} - \omega^2  \omega^{23})+\dot{\alpha}_p  n^{p2})+
$$
$$
+ C^{13}(f^1 \exp{\sigma_0}+ \omega^3 \beta - \omega^2 \gamma)-\dot{\gamma}=0;
$$

\item
$F^0 = 0:$
\begin{equation}\label{70}
\quad C_1 \alpha-C_2 \beta - C_3 \gamma=0.
\end{equation}

For non-solvable groups $C_a=0$.
\end{enumerate}
Equations \eqref{67}-\eqref{70} contain the function $\alpha$:
\begin{equation}\label{71}
\alpha=(\dot{\alpha}_1 +\omega^2 f^3- \omega^3 f^2)\exp\sigma_0.
\end{equation}
We will consider $\alpha$ as an additional function. Then equation \eqref{71}, complementing the system of Maxwell equations, allows us to lower their order. Thus, for the case of Petrov null spaces with unsolvable groups of motions, the system of Maxwell's vacuum equations includes equations \eqref{67}-\eqref{69}, \eqref{71} for the functions:
$$
\alpha, \quad \alpha_a, \quad \sigma_0, \quad \beta,\quad \gamma,\quad \omega^a, \quad \omega^{pq}.
$$
Equations \eqref{67}-\eqref{70} contain the function $\alpha$:
\begin{equation}\label{72}
\alpha=(\dot{\alpha}_1 +\omega^2 f^3- \omega^3 f^2)\exp\sigma_0.
\end{equation}
 The function $\alpha$ will be considered as a new independent function. Then equation \eqref{71}, complementing the system of Maxwell's equations, allows one to lower their order. Thus, for the case of null homogeneous  Petrov  spaces with unsolvable groups of motions, the system of Maxwell vacuum equations includes equations \eqref{67}-\eqref{69}, \eqref{71} for the functions:
$$
\alpha, \quad \alpha_a, \quad \sigma_0, \quad \beta,\quad \gamma,\quad \omega^a, \quad \omega^{pq}.
$$
The functions \quad $\alpha_a, \quad \sigma_0, \quad \omega^a, \quad \omega^{pq} $ \quad
will be called field functions, while the remaining functions are additional (the function $\alpha$ is an example of such an additiona function). During the integration of the system of Maxwell equations, all functions included in each specific solution are divided into two classes. The first class includes field and additional functions, which are arbitrary independent functions of the variable $u^0$. We will call them independent functions. The second class includes field functions that are expressed in terms of independent functions and their derivatives. We will call these functions dependent field functions. The first problem to solve when integrating Maxwell's equations is to find the set of independent functions. After solving the first problem, all remaining dependent field functions must be expressed in terms of independent functions.

\section{Exact solutions of Maxwell equations}

When solving Maxwell equations, two options must be considered separately from each other.

\quad

\noindent
$  \quad 1.\quad {f^2}^2 + {f^3}^2 \ne 0.$

\quad

\noindent
Then the functions $\gamma, \beta$ can be considered as additional functions, and the functions $\omega^{ab}$ can be expressed in terms of them and the functions $f^p = C^{2a}\alpha_a $ as follows:
\begin{equation}\label{73}
\left\{\begin{array}{cc} {\omega^{22}=\frac{{f^2}^2 + \beta^2}{f^2\gamma - f^3 \beta}, \quad \omega^{33}=\frac{{f^3}^2 +\gamma^2}{f^2\gamma - f^3 \beta},}
\\\\{\omega^{23}=\frac{f^{2}f^{3} + \beta\gamma}{f^2\gamma - f^3\beta}.}
\end{array}\right.
\end{equation}
Obviously, in this case, the functions $\omega^{ab}$ become dependent field functions.

\quad

\noindent
$2. \quad f^p=0  \quad  \Rightarrow  \beta = \gamma =0.$

\subsection{Exact solutions of Maxwell equations for the Petrov spaces of type $V_4^{(0)}(VIII)$}

From equations \eqref{21} it follows:
$$
C^{13}= C^{31} =1 \quad C^{22}= -1, \quad f^1=\alpha_3, \quad f^2=-\alpha_2, \quad f^3=\alpha_1.
$$
If \quad $f^3= f^2=0$ \quad ($f^p=0$) \quad from Maxwell equations it follows $\alpha_3=0 \Rightarrow $   the electromagnetic field disappear.
Then ${\alpha_1}^2 +{\alpha_2}^2 \ne 0$, and equations \eqref{72} take the form:
\begin{equation}\label{74}
\left\{\begin{array}{cc} {\omega^{22}= -\frac{{\alpha_2}^2 + \beta^2}{\alpha_2\gamma +\alpha_1\beta},
\quad \omega^{23}= \frac{\alpha_1\alpha_2 - \beta\gamma}{\alpha_2\gamma +\alpha_1\beta},}
\\\\{\omega^{33}=-\frac{\alpha_1^2 + \gamma^2}{\alpha_2\gamma +\alpha_1\beta}}
\end{array}\right.
\end{equation}
Thus, the functions $\omega^{ab}$ become dependent field functions. The functions $\alpha, \beta, \gamma$ become
(independent) additional functions.
Now first part of Maxwell equations can be represented as following:
\begin{equation}\label{75}
\left\{\begin{array}{cc} {\omega^3 \alpha_2 + \omega^2 \alpha_1= \alpha\exp{\sigma_0}-\dot{\alpha}_1,}
\\\\{\omega^3\beta - \omega^2 \gamma = \dot{\gamma}-\alpha_3\exp{\sigma_0} .}
\end{array}\right.
\end{equation}
It follows from this:
\begin{equation}\label{76}
\left\{\begin{array}{cc} {\omega^2 = \frac{(\alpha\beta + \alpha_2\alpha_3)\exp{\sigma_0} - (\dot{\alpha}_1\beta +\alpha_2\dot{\beta})}{\alpha_2 \gamma + \alpha_1 \beta},}
\\\\{\omega^3 = \frac{(\alpha\gamma-\alpha_1\alpha_3)\exp{\sigma_0} + (\alpha_1\dot{\gamma}-\dot{\alpha}_1\gamma)}{\alpha_2 \gamma + \alpha_1 \beta}.}
\end{array}\right.
\end{equation}
We eliminate the function $ \exp{\sigma_0}$ from the system of equations \eqref{75}. As a result, we obtain the equation:
\begin{equation}\label{77}
(\alpha_2 \omega^3 + \alpha_1 \omega^2)\alpha_3 + (\beta \omega^3 - \gamma\omega^2)\alpha= \dot{\gamma}\alpha -\alpha_3\dot{\alpha}_1,
\end{equation}
Remaining part of Maxwell equations can be represented in the form:
\begin{equation}\label{78}
\left\{\begin{array}{cc} {\dot{\beta}+ \alpha \omega^3 +(\dot{\alpha}_2 +\alpha_3 \omega^3)\omega^{23} + (\dot{\alpha}_3 - \alpha_2\omega^2)\omega^{33} + \omega^1 \gamma =0,}
\\\\{- \dot{\alpha}+ \alpha \omega^2  + \alpha_3 (\omega^3 \omega^{22}-\omega^2\omega^{23}) + \omega^1 \beta =0 .}
\end{array}\right.
\end{equation}
By eliminating the function $\omega^1$ from the system of equations \eqref{78} and using equation \eqref{77} we obtain the equation:
$$
\dot{\beta} \beta + \dot{\alpha}\gamma +\dot{\alpha_p \omega^{p3}}\beta + \alpha (\omega^3\beta - \omega^2 \gamma)+ \alpha_3(\omega^3 \alpha_2 + \omega^2 \dot{\alpha}_1)=0 \Rightarrow
$$
\begin{equation}\label{79}
(\frac{1}{2}\beta^2 + \alpha\gamma +\alpha_3\beta \omega^{33})_{,0} =  \alpha_3(\gamma \omega^{23})_{,0}- \beta \omega^{23}\dot{\alpha}_2.
\end{equation}
Equation \eqref{79} is the key to solving Maxwell equations.
All functions $\alpha,  \alpha_a, \beta, \gamma$  and their derivatives with respect to $u^0$ in the equation \eqref{79} are bound. To unbind them one can introduce the additional function:
\begin{equation}\label{80}
\rho = \frac{1}{2}\beta^2 + \alpha\gamma +\alpha_3\beta \omega^{33}.
\end{equation}
Equation \eqref{77} takes the form:
\begin{equation}\label{81}
\dot{\rho } =\alpha_3(\gamma \omega^{23})_{,0}- \beta \omega^{23}\dot{\alpha}_2.
\end{equation}
Let us find all solutions to the system of equations \eqref{80} \eqref{81}. Substituting these solutions into the functions $\omega_a$ and $\omega^{ab}$, we obtain all nonequivalent solutions of Maxwell's equations

If \quad $\gamma \omega^{23}\ne const, $ \quad the solution of the equation \eqref{79} can be represented in the form:
$$
\alpha_3=\frac{1}{(\gamma\omega^{23})_{,0}}(\dot{\rho } + \beta \omega^{23}\dot{\alpha}_2), \quad \alpha = \frac{1}{\gamma}(\rho -\frac{1}{2}\beta^2-\alpha_3 \beta \omega^{33}).
$$
The dependent functions $\omega^a, \omega^{pq}$ are expressed from the relations \eqref{74}, \eqref{76}, \eqref{78} in terms of the independent functions. In present solution, the independent functions are the functions $\alpha_1,$ $ \alpha_2,$ $ \beta, $ $ \gamma, $ $ \rho,$ $ \sigma_0$.
\quad

Let \quad $\gamma \omega^{23}= c \quad $ (from here $ \xi =\pm 1 $). One has to consider three options:

\begin{enumerate}

\item
$ c\ne 0.$
The system of equations \eqref{79} \eqref{80} has the form:
\begin{equation}\label{82}
\left\{\begin{array}{cc}{\rho =\frac{1}{2}\beta^2 + \alpha\gamma +\alpha_3\beta \omega^{33}, \quad  \gamma\dot{\rho} +c\beta\dot{\alpha}_2=0,} \\{\gamma\alpha_2(\alpha_1- c)-\beta(\gamma^2 +c\alpha_1) =0.}
\end{array}\right.
\end{equation}
Let us enumerate all solutions of the system of equations \eqref{82}.

\begin{enumerate}

\item $\beta \ne 0.$  From the set of equations \eqref{80} -  \eqref{82} it follows:
\begin{equation}\label{83}
\alpha_2 = a, \quad \alpha = b -\frac{\beta^2}{2} +\alpha_3 \beta \frac{{\alpha_1}^2 +\gamma^2}{a\gamma +\alpha_1\beta}.
\end{equation}
$$
\gamma= \frac{1}{2\beta}(\alpha_2(\alpha_1 - c) + \xi\sqrt{a^2(\alpha_1-c)^2-4c\alpha_1 \beta^2}).
$$
The dependent functions are $\alpha, \alpha_2, \gamma,$ and also the functions $\omega^a, \omega^{pq}$. The independent functions are $\alpha_1,$ $ \alpha_3,$ $ \beta, $ $ \sigma_0$.

\quad

\item $\beta=0 $

From the set of equations \eqref{82} it follows:
$$
 \rho =b, \quad \alpha = \frac{b}{\gamma}, \quad \alpha_1= c.
$$
Obviously, \quad $\alpha_2\ne 0$ \quad otherwise \quad $\omega^{22}=0$, \quad which violates the Lorentz signature of the space. The functions $\omega^a, \omega^{pq}$ are found from the relations \eqref{74}, \eqref{76}, \eqref{78} and have the following form:
$$
 \omega^{22} =-\frac{\alpha_2}{\gamma}, \quad \omega^{23} = \frac{c}{\gamma}, \quad \omega^{33} = -\frac{c^2 + \gamma^2}{\alpha_2\gamma},
$$
$$
\omega^2 =\frac{\alpha_3 \exp\sigma_0}{\gamma},\quad \omega^3 =\frac{(b-\alpha_3) \exp\sigma_0 + c\dot{\gamma}}{\alpha\gamma}.
$$
$$
\omega^1 =\frac{1}{\alpha_2\gamma^3}((c^2{\alpha_3}^2 -b^2 - \alpha_2\alpha_3(c^2 + \gamma^2))\exp\sigma_0- $$
$$-c(b\dot{\gamma}+\gamma\dot{\alpha_2}\alpha_2)+\gamma\dot{\alpha}_3(c^2 + \gamma^2)).
$$

\end{enumerate}

\item $\gamma =0.$ The solution of the system of equations \eqref{79} \eqref{80} is:
$$
\alpha_3 = \frac{1}{2\alpha_1}(\beta^2+{\alpha_2}^2 + b).
$$
The functions $\omega^a, \omega^{pq}$ are found from the relations \eqref{74}, \eqref{76}, \eqref{78} through the independent functions $\alpha_1,$ $\alpha_2,$ $ \alpha,$ $ \sigma_0$ and have the following form:
$$\omega^{23} = \frac{\alpha_2}{\beta}, \quad \omega^{33} = -\frac{\alpha_1}{\beta}, \quad \omega^{22}=-\frac{{\alpha_2}^2+ \beta^2}{\alpha_1 \beta}
$$
$$
\omega^2 =\frac{(\alpha\beta+\alpha_2\alpha_3)\exp\sigma_0 -\dot{\alpha}_1\beta}{\alpha_1\beta},\quad \omega^3 =-\frac{\alpha_3 \exp\sigma_0}{\gamma}.
$$
$$
\omega^1= \frac{\dot{\alpha} - \alpha \omega^2  + \alpha_3 (\omega^2\omega^{23}-\omega^3 \omega^{22})}{ \beta } .
$$

\item  $\omega^{23}=0$. From this condition  it follows:
$$
\beta =\frac{\alpha_1\alpha_2}{\gamma} \Rightarrow \omega^{33}= -\frac{\gamma}{\alpha_2}, \quad \omega^{22}= -\frac{\alpha_2}{\gamma },
$$
The solution of the system of equations \eqref{79} \eqref{80} has the form:
$$
\alpha = \frac{1}{2\gamma^3}((2\alpha_1\alpha_3 + b)\gamma^2 -(\alpha_1\alpha_2)^2), \quad \rho  =b.
$$
The functions $\omega^a$ are expressed through independent functions $\alpha_a, \gamma, \sigma_0$ from the relations \eqref{76}, \eqref{78}.

\end{enumerate}

\quad

\subsection{Exact solutions of Maxwell equations for the spaces of types $V_4^{(-)}(VIII),  V_4^{(+)}(VIII), V_4(IX)$}

Maxwell equations \eqref{67} -- \eqref{70} for homogeneous Petrov spaces of type $V^{*}_4(N)$ differ from each other only in the structure constants $C^c_{ab}$ (or $C^{ab}$).
As follows from equations \eqref{23}, \eqref{25}, \eqref{32}, the structure constants for spaces of type
\begin{equation}\label{91}
V^{*(-)}_4(VIII), \quad V^{*(-)}_4(VIII), \quad V^{*}_4(IX)
\end{equation}
can be represented as:
\begin{equation}\label{85}
C^{ab}=\varepsilon_a\delta^{ab}  \Rightarrow f^a = \varepsilon_a \alpha_a.
\end{equation}
Values $\varepsilon_a$ have the form:
\begin{enumerate}
\item \quad  $V^{(-)}_4(VIII), \quad \varepsilon_1 =-1, \quad \varepsilon_2 = \varepsilon_3 =1,$

\item \quad  $V^{(+)}_4(VIII), \quad \varepsilon_1 =1\quad \varepsilon_2 =-1 \quad \varepsilon_3 =1,$

\item \quad  $ V_4(IX), \quad \varepsilon_a =1. $

\end{enumerate}
Substituting the expressions \eqref{85} into Maxwell's equations yields the following systems of equations:
\begin{equation}\label{86}
\left\{\begin{array}{cc} {\omega^3 \beta - \omega^2 \gamma=-\varepsilon_1(\dot{\alpha}+\alpha_1\exp{\sigma_0}),}
\\\\{\omega^3 \alpha_2 \varepsilon_2 -  \omega^2 \alpha_3 \varepsilon_3 =(\dot{\alpha}_1 - \alpha\exp{\sigma_0});}
\end{array}\right.
\end{equation}
\begin{equation}\label{87}
\left\{\begin{array}{cc} {\alpha \omega^3 +\dot{\alpha}_p \omega^{p3} +\varepsilon_1 \alpha_1(\omega^3 \omega^{23} - \omega^2 \omega^{33}) +\omega^1 \gamma - \varepsilon_2\dot{\beta}=0,}
\\\\{ \alpha \omega^2 +\dot{\alpha}_p \omega^{p2}+ \varepsilon_1 \alpha_1(\omega^3 \omega^{22} - \omega^2 \omega^{23}) +\omega^1 \beta + \varepsilon_3\dot{\gamma}=0.}
\end{array}\right.
\end{equation}
Let us eliminate the function $\omega^1$ from the system of equations \eqref{87}. As a result, we obtain the equation:
\begin{equation}\label{88}
\alpha(\beta \omega^3 - \gamma \omega^2)+(\dot{\alpha}_2 +\varepsilon_1 \alpha_1 \omega^3) (\beta \omega^{23} - \gamma \omega^{22}) +
\end{equation}
$$
+(\dot{\alpha}_3 -\varepsilon_1 \alpha_1 \omega^2)(\beta \omega^{33} - \gamma \omega^{23})
=(\dot{\beta}\beta \varepsilon_2  + \dot{\gamma}\gamma \varepsilon_3);
$$
 We eliminate the function $\exp \sigma_0$ from the system of equations 87\eqref{86}. As a result, we obtain the equation:
$$
\varepsilon_1\alpha(\beta \omega^3 -\gamma \omega^2) - \alpha_1(\varepsilon_2 \alpha_2 \omega^3 -\varepsilon_3\alpha_3\omega^2)= -(\alpha^2 \dot{\alpha}^2+{\alpha_1}\dot{\alpha^1}).
$$
 Using this equations one can transforms the equations \eqref{88} to the form:
\begin{equation}\label{89}
\varepsilon_1(\alpha^2 + \alpha_1^2)+ \varepsilon_2(\beta^2 + \alpha_2^2) + \varepsilon_3(\gamma^2 + \alpha_3^2) = c=const.
\end{equation}
Let's consider all options. The option ${\alpha_2}^2 + {\alpha_3}^2 \ne 0 $ we consider first.
\begin{enumerate}

\item ${\alpha_2}^2 + {\alpha_3}^2 \ne 0. $

The conditions \eqref{73} have the form:
\begin{equation}\label{90}
\left\{\begin{array}{cc} {\omega^{22}= \frac{{\alpha_2}^2 + \beta^2}{\varepsilon_2\alpha_2\gamma -\varepsilon_3\alpha_3\beta}, \quad \omega^{23}= \frac{\varepsilon_2\varepsilon_3\alpha_2\alpha_3 + \beta\gamma}{\varepsilon_2\alpha_2\gamma -\varepsilon_3\alpha_3\beta},}
\\\\{\omega^{33}=\frac{\alpha_3^2 + \gamma^2}{\varepsilon_2\alpha_2\gamma -\varepsilon_3\alpha_3\beta}}
\end{array}\right.
\end{equation}
From equations \eqref{76} it follows:
\begin{equation}\label{91}
\omega^2 = \frac{\left( (\varepsilon_1\varepsilon_2\alpha_1\alpha_2  -\alpha\beta)\exp{\sigma_0} + \dot{\alpha}_1\beta +\varepsilon_1\varepsilon_2\dot{\alpha} \alpha_2 \right)}{\varepsilon_2\alpha_2\gamma -\varepsilon_3\alpha_3\beta},
\end{equation}
$$
\omega^3 = \frac{\left( (\varepsilon_1\varepsilon_3\alpha_1\alpha_3  -\alpha\gamma)\exp{\sigma_0} + \dot{\alpha}_1\gamma +\varepsilon_1\varepsilon_3\dot{\alpha} \alpha_3  \right)}{\varepsilon_2\alpha_2\gamma -\varepsilon_3\alpha_3\beta}.
$$
Function $\omega^1$ can be found from the first (if $\gamma \ne 0$) or from the second (if $\gamma=0$) equation of the system of equations  \eqref{87}. Let us consider possible options.

\begin{enumerate}

\item $\gamma \ne 0.$

In this case we use the first equation from the system \eqref{87}:
\begin{equation}\label{92}
\omega^1 =\frac{1}{\gamma(\varepsilon_2\alpha_2\gamma -\varepsilon_3\alpha_3\beta)}
((\varepsilon_2\alpha_2 \gamma -\varepsilon_3\alpha_3\beta)(\varepsilon_2\dot{\beta}-\alpha \omega^3)+
 \end{equation}
$$+(\varepsilon_2\varepsilon_3\alpha_2\alpha_3 + \beta\gamma)(\varepsilon_1\alpha_1 \omega^3 +\dot{\alpha}_2)
+(\dot{\alpha}_3-\varepsilon_1\alpha_1 \omega^2 )({\alpha_3}^2 + \gamma^2) ).$$

\quad

\quad

\item $\gamma=0.$ In this case we use the second equation from the system \eqref{87}:
\begin{equation}\label{93}
\omega^1 =\frac{1}{\beta^2\alpha_3}
(\varepsilon_2\alpha_2\alpha_3(\dot{\alpha}_3-\varepsilon_1\alpha_1\omega^2) +\varepsilon_3(\dot{\alpha}_2+\varepsilon_1\alpha_1\omega^3)({\alpha_2}^2+\beta^2)-
\alpha\alpha_3\beta \omega^2).\end{equation}

\end{enumerate}

\item $\alpha_2 = \alpha_3 =0.$

From equations \eqref{87} - \eqref{90} it follows:
\begin{equation}\label{94}
\alpha=c\cos{\tau}, \quad  \alpha_1 = c\sin{\tau},  \quad \sigma_0 = \ln \dot{\tau}.
\end{equation}
Let us introduce conditions \eqref{94} in system  of equations \eqref{87}. In result we obtain the system:
\begin{equation}\label{95}
\left\{\begin{array}{cc} {\omega^3 \omega^{23} - \omega^2 \omega^{33} =0,}
\\\\{\omega^3 \omega^{22} - \omega^2 \omega^{23} =0}
\end{array}\right.\Rightarrow
\end{equation}
$\omega^2=\omega^3 =0,$  because  $(det|\omega^{pq}| \ne 0)$.  Thus the solution is:
$$
\alpha_1 = \sin {\tau}, \quad \sigma_0 = \ln{\dot{\tau}}, \quad \omega^p=\alpha_p = 0, \quad \omega^{22} =\frac{{\omega^{23}}^2 +1}{\omega^{33}},
$$
$\tau, \quad \omega^1, \quad \omega^{23}, \quad \omega^{33}\quad $ are arbitrary functions on $u^0.$

\end{enumerate}

\quad

\quad

\section{Conclusion}

Unlike non-null homogeneous Petrov spaces, the classification of exact solutions to the field equations for null homogeneous Petrov spaces has not yet been considered. Note that a complete classification of exact solutions of the vacuum Maxwell equations in non-null homogeneous Petrov spaces is presented in \cite{57}, \cite{58}. The problem of classifying vacuum and electrovacuum non-null homogeneous Petrov spaces of type $V_4(I)$ was completely solved in \cite{60}, \cite{61}. This paper presents the first example of classifying exact solutions for null homogeneous Petrov spaces. The classification is carried out for Maxwell equations in null homogeneous Petrov spaces with unsolvable groups of motions. To finally resolve the problem of classifying exact solutions to the vacuum Maxwell equations in homogeneous Petrov spaces, it remains to consider the case with solvable groups of motions. This will be done in a separate article (in preparation).

\quad

Data availability
No data was used for the research described in the article.

\quad

Conflictsofinterest{{ The authors declare no conflicts of interest } 
}

\quad

\end{document}